# Measuring Information Exchange and Brokerage Capacity of Healthcare Teams

Grippa, F., Bucuvalas, J., Booth, A., Alessandrini, E., Fronzetti Colladon, A., & Wade, L. M.

# Measuring Information Exchange and Brokerage Capacity of Healthcare Teams

Grippa, F., Bucuvalas, J., Booth, A., Alessandrini, E., Fronzetti Colladon, A., & Wade, L. M.

**Abstract**

**Purpose** – In this paper we explore possible factors impacting team performance in healthcare, by focusing on information exchange within and across hospital's boundaries.

**Design/methodology/approach** – Through a web-survey and group interviews, we collected data on the communication networks of 31 members of four interdisciplinary healthcare teams involved in a system redesign initiative within a large US children's hospital. We mapped their internal and external social networks based on management advice, technical support, and knowledge dissemination within and across departments, studying interaction patterns that involved more than 700 actors. We then compared team performance and social network metrics such as degree, closeness and betweenness centrality, and computed cross ties and constraint levels for each team.

**Findings** – The results indicate that highly effective teams were more inwardly focused and less connected to outside members. Moreover, highly recognized teams communicated frequently but, overall, less intensely than the others.

**Originality/value** – Mapping knowledge flows and balancing internal focus and outward connectivity of interdisciplinary teams may help healthcare decision-makers in their attempt to achieve high value for patients, families and employees.

## Introduction

As recently highlighted in literature, the health care sector is an environment that is rich in isolated silos and professional "tribes" in need of connectivity" (Long, Cunningham and Braithwaite, 2013; Sexton *et al.*, 2017). The healthcare community is increasingly recognizing the need to find new approaches to improve both outcomes and the overall experience for patients and healthcare workers. It has been widely demonstrated that the majority of the avoidable adverse events are due to the lack of effective communication and collaboration, with an estimated 80 percent of serious medical errors involving miscommunication during the hand-off between medical providers (Solet *et al.*, 2005). Defining clear handoff practices, reducing interruptions and distractions, ensuring a common understanding about the patient and clarification of transition of responsibility are all key factors to reduce errors and improve patient safety (Palmieri, P.A., DeLucia, P.R., Oh, T.E., Peterson, L.T., & Green, 2008). As reported by several studies over the past two decades (Kohn, Corrigan and Donaldson, 2000; Landrigan *et al.*, 2010; Makary and Daniel, 2016), medical error is the third leading cause of death in the US. A recent literature review (James, 2013) described an incidence range of 210,000-400,000 deaths a year associated with medical errors among hospital patients. Most of the errors are often related to issues found in the health care system rather than problems attributable to individual errors.

One of the approaches to solve this problem is to improve the communication processes within and across hospital units, building interdisciplinary teams to help reduce the multiple gaps that exist among professions, departments, and specialties including the clinician-patient divide (Awad *et al.*, 2005; Long, Cunningham and Braithwaite, 2013). Working in teams has been demonstrated to reduce errors as medical staff rely on each other's expertise and specialized knowledge (Dutton *et al.*, 2003; Chin *et al.*, 2004; Lemieux-Charles and McGuire, 2006).

By creating interdisciplinary and cross-functional healthcare teams, hospitals have the opportunity to balance the trade-off between exploitation (internal focus) and exploration (external focus), creating the foundations for a true ambidextrous organization (Tushman, 1997; O'reilly and Tushman, 2004). This requires selecting individuals with the right combination of skills, hierarchical position, status and external connections, which can affect the exploration and exploitation of new knowledge and impact the trade-off in team composition (Perretti and Negro, 2006).

This case study describes how healthcare teams exchange information within and across boundaries, search for new knowledge in order to create a completely new care delivery system and in doing so, rely on internal ties and knowledge of the process. The healthcare teams involved in this study were composed of professionals involved in making patient/medical decisions (e.g. nurses, physicians) as well as by others whose decisions impact health outcomes and safety (e.g. director of patient/family experience, head of the ER unit).

## Literature Review

The present study is based on the recognition that teams are an essential component for bridging the gaps between isolated units within hospitals. Our case study relies on a definition of teams as complex systems made of individuals "who are interdependent in their tasks, who share responsibility for outcomes, who see themselves and who are seen by others as an intact social entity embedded in one or more larger social systems" and who manage their relationships across organizational boundaries (Cohen and Bailey, 1997, p. 241). Our focus is on the task-related team, defined as a group of individuals whose task requires members to work together to produce something for which they are collectively accountable and whose acceptability is potentially assessable (Hackman, 2004).

Healthcare teams are usually described based on the type of tasks they perform (Lemieux-Charles and McGuire, 2006). Project teams, management teams, and care delivery teams might be distinguished based on their daily activities, which can involve a combination of direct care of patients and designing new health delivery modes. Nevertheless, their actions have a similar impact on patient safety. Each member brings his or her special knowledge and capabilities, but also interpersonal relationships with the members inside and outside of the team (Ancona, Bresman and Caldwell, 2009). Yet even though individual team members may have distinct and complementary expertise, effective teams require close ties among the members, ability to effectively communicate, and organizational support.

In their literature review of health care team effectiveness from 1985 to 2004, Lemieux-Charles and McGuire (Lemieux-Charles and McGuire, 2006) linked outcomes to team effectiveness and to processes like effective team communication and cohesion. They observed that increased team autonomy correlated with decreased hospital readmissions and with higher levels of staff satisfaction and retention. High-functioning teams have been characterized by positive communication patterns and high levels of collaboration and participation (Shortell, 2004; Temkin-Greener *et al.*, 2004). Other studies found that increased team diversity and interdependence are associated with decreased length of stay and hospital charges (Dutton *et al.*, 2003). Further evidence indicated that team communication and training in the use of quality improvement methods was linked with improved patient outcomes.

Studies conducted in other industries found that structurally diverse work groups are characterized by members who use their different organizational affiliations, roles, or positions to expose the team to unique sources of knowledge, which is beneficial for performance (Burt, 2004; Cummings, 2004). In particular, Cummings (2004) found that effective work groups engage in external knowledge sharing, through the exchange of information, know-how, and feedback with important stakeholders outside of the group.

In a study that investigated the association between team constraint and team performance of 15 process improvement teams, Rosenthal (1997) noticed how differences in social networks explain performance variation: teams composed of members with more entrepreneurial networks were more likely to be recognized for improving the quality of plant operations. In a study of 120 new-product development projects undertaken by 41 divisions, Hansen (1999) found evidence that weak inter-department ties help a project team search for knowledge in other departments but impede the transfer of complex knowledge, which relies on strong ties between the two parties to a transfer. Burt (1992, 2005) used social network indicators and performance data from managerial networks across industries (not including healthcare) to demonstrate that networks that span structural holes are associated with creativity and learning, more positive evaluations and more successful teams. Burt (2004) also found that dense networks do not necessarily enhance performance and could be associated with substandard results.

The analysis of collaboration and communication among healthcare staff is a key component of any system redesign initiative that aims at improving quality of care (Wagner *et al.*, 2001; Shanafelt *et al.*, 2010, 2015; Bodenheimer and Sinsky, 2014). While best outcomes depend on productive interactions and communication among members of interdisciplinary healthcare teams, coordination becomes difficult as teams grow in size. In the setting of complex care, teams must gather information from multiple subspecialists, synthesize the information acquired, come to decisions and execute a plan (Delva, Jamieson and Lemieux, 2008; Harrod *et al.*, 2016).

## Method

Complex care often involves input from and coordination with other departments, so information must flow beyond unit, divisional, and departmental boundaries. Reporting relationships increase complexity since team members may belong to distinct departments and many individuals belong to multiple teams. The visualization of these relationships is the first step to recognize interdependencies and bottlenecks. For this reason, in this study we use Social Network Analysis to build social maps and extract centrality indicators that can reveal blockages in the information flows and offer ideas on how to improve team effectiveness.

### Participants

The study participants were 42 employees of a large children's hospital in the US (20 women and 22 men). There was an equal representation of different roles across hospital's units, including physicians, nurses, business directors, AVP and VP of Finance, directors of quality improvement initiatives, clinical pharmacists, anesthesiologists and program managers. Participation in this study was on a voluntary basis. Almost all the hospital units were represented in our sample, with at least one representative for each department. The hospital units that had more than three members participating in the study were: Anesthesia unit, Health Improvement Center, Patient Services and Patient/Family Experience, Gastroenterology and the Heart Institute. Participants were not working together at the time of the study. They represented different units and departments that were also located in geographically distant hospital campuses. Each members was assigned to a team based on work experience, functional unit, and tenure within the organization. For example, the Director of Finance and the AVP of Finance were both assigned to the team in charge of learning how the hospital costs were affecting value creation for patients, families and employees. We recognize that individual differences, tenure within the organization and knowledge of the topic could impact the team outcomes. Each team was composed of both senior and junior employees with a tracked record of expertise in their respective area. Members with experience in other service industries were also included.

**Instruments**

Through a web-questionnaire sent via email we asked participants to report up to 25 people within and outside the hospital they would go to when: 1) looking for advice based on subject matter expertise; 2) seeking support for their career development; 3) seeking technical support; or 4) sharing new ideas. Out of the 42 team members involved in the project, 31 responded to the survey (72% response rate).

Using the name generator technique (Burt *et al.*, 2012) we created a list of 700 unique contacts with whom respondents communicated more frequently within and outside the hospital. This allowed the creation of four different social networks, based on connections among individuals seeking managerial advice, sharing new ideas, looking for an expert opinion during complex cases, and for solving technical problems, both within and across the hospital's boundaries. To map and measure the internal and external social networks we used metrics of Social Network Analysis (Burt, 1992; Wasserman and Faust, 1994; Cross, Prusak and Parker, 2002), which helped to identify brokers, boundary spanners and central connectors who can transfer knowledge between departments and increase collaboration.

| METRIC | DEFINITION |
|---|---|
| Degree Centrality | The total number of ties a node has to other nodes. |
| In-degree Centrality | Number of incoming ties, representing received requests of advice, knowledge sharing and technical support |
| Out-degree Centrality | Number of outgoing ties, representing requests of advice, knowledge sharing and technical support made by each individual. |
| Closeness Centrality | The average length of the paths linking a node to all others. This measure can sometimes be seen as a proxy of the speed with which a node can be reached, or can reach the others(Wasserman and Faust, 1994). |
| Betweenness Centrality | The extent to which a node is connected to other nodes that are not connected to each other. It is a measure of the degree to which a node serves as a bridge, mediating, for instance, a request of advice. |
| Network Constraint | Measures the extent to which an actor's network is a limitation around him/her, limiting his or her vision of alternative ideas and sources of support. Network constraint is an index that measures the extent to which a person's contacts are also linked among themselves, closing the triads (i.e., if A is connected to B and C, there is also a link between B and C). A social actor who can mediate a connection between unlinked peers, can take advantage of his/her social position and choose, for example, a 'divide et impera' strategy, or be the broker of good ideas (Burt, 2004). In this example, we have a 'structural hole', which is the missing link between B and C, and therefore a lower value of network constraint for A. |
| Cross Ties | Number of links towards actors belonging to social clusters different from theirs. |

**Table 1.** Metrics of Social Network Analysis used in the study

**Procedure**

Participants were assigned to five teams, whose goal was to conduct a preliminary inventory of strengths, weaknesses and opportunities to improve the current system of care delivery at the hospital, and learn how the organization impacted the experience of their patients, families and staff. The teams were charged with finding exemplars in value delivery, both in healthcare, and other industries. They were prompted to look outside the hospital boundaries at organizations in other industry that had excelled in quality of service and personalization of the experience (e.g. The Walt Disney Company). The final deliverable was an assessment of the current situation of the hospital with a proposal of improvements with regards to five areas: safety, patient and family experience (PFE), employee engagement and team function (EETF), healthcare outcomes and costs. The five teams were charged with exploring challenges and opportunities of a new care delivery system that could result in a quantum leap in improvement of outcomes, patient/family and provider experience. The team members worked together over a period of six months and presented their findings during a two-day synthesis session. They generated extensive reports to describe the status quo for the five subjects and offered recommendations for improvements. Members of the teams met face-to-face during bi-weekly meetings to engage in design, prototyping, testing and implementation of a new healthcare delivery system.

The team focused on "Safety" was excluded from the analysis since their report/deliverable was missing at the time of the observation. The analysis included four teams, plus an operational team whose members coordinated their work to guarantee a seamless process.

In order to understand the mechanisms that could lead to effective teamwork in healthcare, we collected different variables. Team performance was assessed by team leaders at the end of the six month-period, and was operationalized based on the number and quality of insights as well as their impact on the project. Team leaders were asked to assess the teams on three criteria: originality of the findings; number of findings; impact of findings on the overall project in terms of quality and usefulness. Scores spanned from 0 (=very low) to 5 (=extremely high).

To understand the degree of internal connectivity of teams (Wasserman and Faust, 1994; Everett and Borgatti, 2005), we used social network analysis metrics that can offer insights on the internal dynamics and existing ties among members (Cummings, 2004). These metrics are described in Table 1 and include: degree centrality, in-degree centrality, out-degree centrality, closeness centrality. To identify the ability of team to be outwardly connected, we selected betweenness centrality, network constraints and cross ties. These metrics offer the opportunity to measure the brokerage capacity of team members to establish connections with other units and teams. Indicators of brokerage capacity measure the average ability of team members to serve as bridges within or outside their team, while connectivity focuses on the direct contacts of team members, in terms of number of incoming and outgoing ties, as well as the degree to which a team member is near all other members and therefore more embedded at the network core. Prior studies have highlighted the benefits of key social network positions in networks such as advice and trust (Battistoni and Fronzetti Colladon, 2014). By observing both connectivity and outward connectivity, we aim at measuring the members' ability to explore new, radical ideas coming from other industry and other departments, and to exploit the already existing knowledge within the organization (O'reilly and Tushman, 2004). As Hargadon suggested (2005, p. 17), by holding a central position in their informal social networks, individuals are more likely "*to acquire knowledge without acquiring the ties that typically bind such knowledge to particular worlds*".

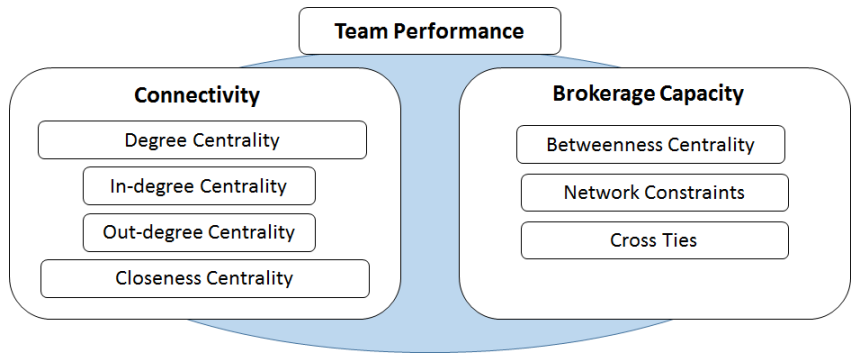

**Figure 1.** Variables represented in the study

## Results

To visually represent how frequently members cross the organizational boundaries to access critical information, we mapped information flows among the departments and among team members. Figure 2 identifies the teams whose members potentially acted as knowledge brokers, showing a lower "network constraint" score, who were in a position to better facilitate the exchange of information across hospital units and teams. Actors with low constraints have more opportunities for brokering, as well as an advantage with respect to information access (Burt, 1992). Most of the knowledge brokers were members of the Outcomes Team, spanning connections across different departments and outside stakeholders. Operational Team members, who coordinated the entire improvement project, and members of the PFE and EETF teams were deeply embedded in multiple workgroups, playing various roles across departments and acting as ambassadors of the project.

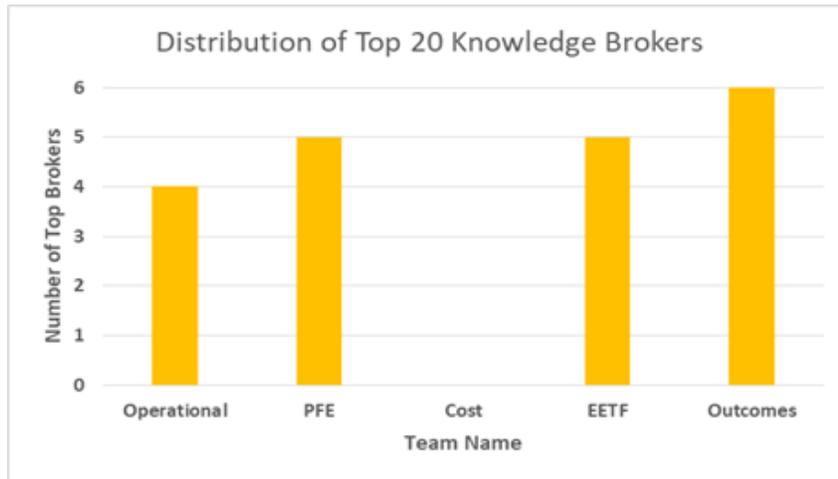

**Figure 2**. Top knowledge brokers across teams (EETF stands for Employee Engagement and Team Function, and PFE stands for Patient-Family Experience).

Figure 3 illustrates the variation in out-group communication for each team. The Outcome team and the Patient and Family Experience team (PFE) had more ties with external stakeholders than the Cost and EETF teams.

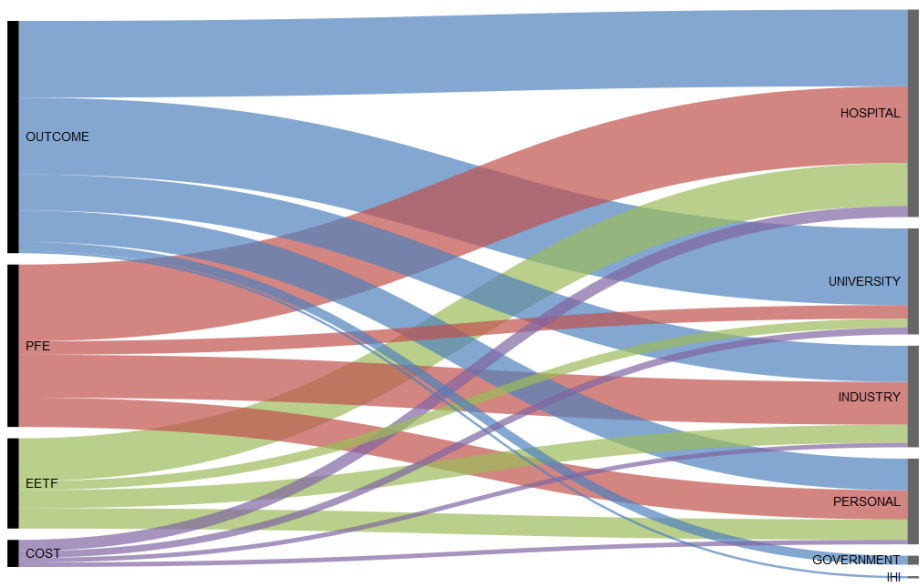

**Figure 3.** Out-group communications.

In general, teams had more external contacts with other hospitals, or university departments. Other external links were with people working in the healthcare industry (private companies), personal contacts or employees of the government or of the Institute for Healthcare Improvement (IHI). The Outcome team had more heterogeneous contacts. The PFE team also had a significant amount of communication which cross the organizational boundaries. Cost and EETF teams, on the other hand, had significantly lower interactions with potential external knowledge sources.

Figure 4 reports the metrics of social interaction for each team by differentiating between out-group and in-group metrics, as well as between their brokerage capacity and network connectivity. For the in-group and out-group communications, the PFE and Outcome teams have more cross ties and higher betweenness centrality, which indicate a stronger effort to connect across boundaries and tap into other units' expertise (i.e. a higher brokerage capacity). With respect to connectivity, we see that Outcome and PFE teams have more outgoing ties and are closer to the network core. PFE also shows high values of indegree centrality, proving its significant amount of communication also within the hospital boundaries. In addition, PFE and Outcome teams are more central with respect to closeness, we see how, overall, they outperform the Cost and EETF teams in terms of connectivity.

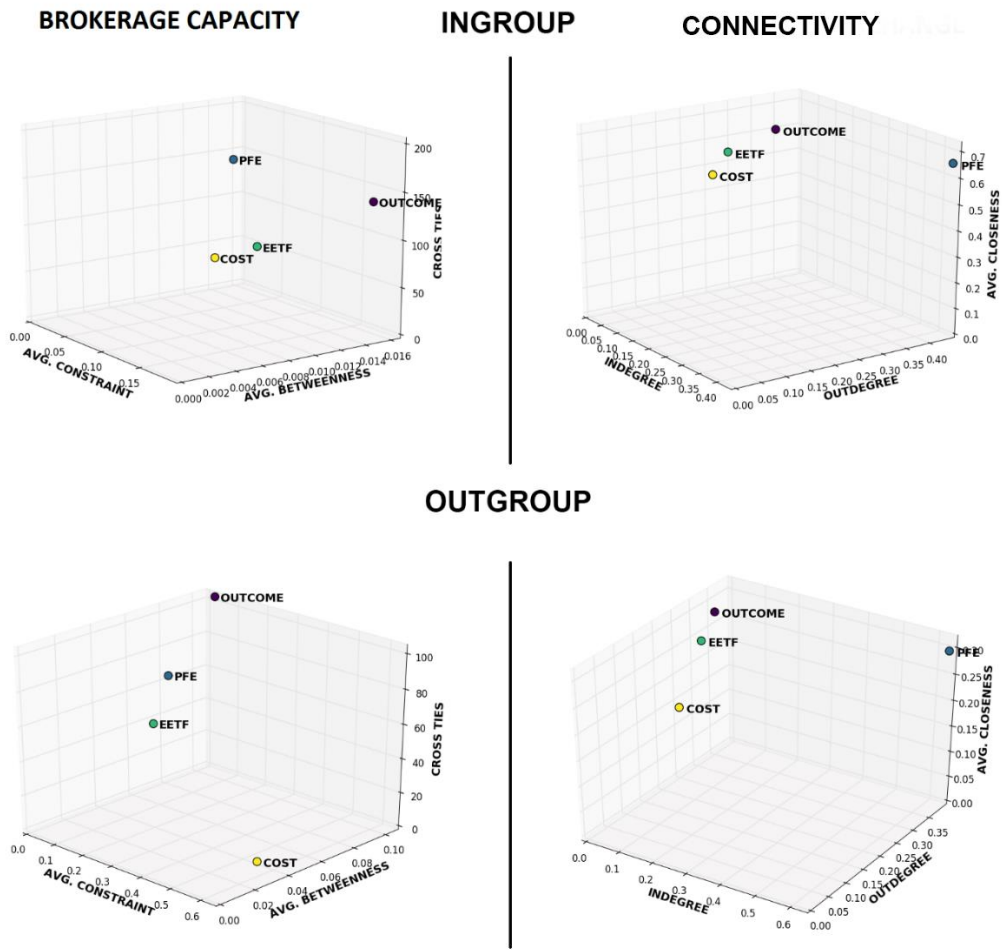

**Figure 4.** Team Position based on Connectivity and Brokerage Capacity metrics.

Figure 5 illustrates the scores associated to the work of the four teams based on an assessment of number of findings, originality/quality of findings and impact. The Employee Engagement and Team Function and the Cost Teams received the highest scores in all the three criteria.

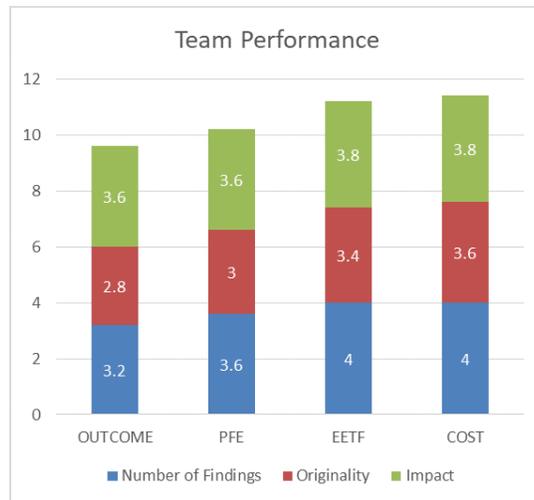

**Figure 5.** Team performance based on number, quality and impact of findings.

In summary, our findings indicate that teams that perform better have an inverse relationships with brokerage capacity and connectivity. They have less frequent interactions with external knowledge sources and are less embedded also in the internal network, which translates in a lower closeness and less direct connections. On the other hand, teams whose report received lower scores (PFE and Outcomes) were highly connected with other hospital units, government agencies and industry professionals. The findings seem to suggest that highly ranked teams are focused more inwardly, and their members are less central with respect to the full network (lower values of average betweenness and closeness centrality). Members of the Cost team and EETF team (for the out-group) had higher network constraint scores (Burt, 2004), indicating a higher closure of their ego-networks, which indicates that each of the member's contacts is connected to his/her other contacts. This means that closer relationships with their team members – with a lower number of direct contacts to manage and less brokerage ties – produced more effective knowledge sharing and efficient communication processes.

## Discussion

This study confirms previous research on team effectiveness (Gupta, Smith and Shalley, 2006; Siggelkow and Rivkin, 2006) describing the relationship between team performance and communication as having an inversely u-shaped form: team effectiveness can be pursued by balancing exploitation (internal focus) and exploration (external focus) and by avoiding excessive or inadequate communication. In this study we found that highly effective teams were more inwardly focused and less connected to outside members. Members of the out-group were both employees working in other hospital units and individuals outside the hospital connected to the participants.

The results indicate that teams who scored the highest in terms of quality, originality and impact of findings (EETF and Cost) communicated frequently but, overall, less intensely than others. Having less scattered communication seems to be associated with higher team effectiveness, as teams may focus on the immediate deliverable and have more efficient conversations. This seems to be aligned with other studies showing that a dense network does not necessarily enhance performance and could be associated with substandard results (Burt, 2004). We find that acting as broker and facilitating information flows might not always conduce to higher recognition. Consistently, our results seem to suggest that a large number of cross-ties between team members and people outside the hospital is not necessarily associated to increased team performance. While innovation has been associated in the past with the ability of teams and organizations to cross institutional boundaries and tap into new ideas and different perspectives (Ancona, Bresman and Caldwell, 2009), there is a fine balance between excessive communication and inadequate interaction with various stakeholders. A not too high level of inter-group connections is more likely to lead to the highest performance "*by enabling superior ideas to diffuse across groups without reducing organizational diversity too quickly*" (Fang, Lee and Schilling, 2010, p. 625).

Our exploratory study found that higher network constraint levels are possibly conducive to higher team performance. Brokers have a very important role in the long term especially when the project becomes increasingly oriented to outside stakeholders, rather than toward internal team operations. The brokers identified in each team might become strategic partners, or champions, when the project enters the next phase, where their network position will facilitate the creation of interfaces with other external organizations and outside members. An explanation for our finding is that the teams had a limited time to get to know each other, understand how every member could contribute to the overall goal, and leverage each other's knowledge, both tacit and explicit. Teams who produced a more impactful deliverable had a more focused communication, dispersed over a lower number of connections and with fewer cross ties with external stakeholders. This might have helped members to stay focused and leverage each other's informal connections within the team. It is important to remember that the final deliverable was a report containing information and suggestions on the current state of the hospital with regard to healthcare cost, healthcare outcomes, employee engagement and patient experience. It could be that teams with fewer external ties had a better chance to focus on collecting relevant institutional knowledge, while others who had higher external ties might have been pulled into different directions and could have been less focused on their task. In particular, the EETF team was composed of members who had immediate access to internal knowledge repository and a direct formal ties to the HR department, which helped locate the right information in the most efficient way. Because of the strong ties of the EETF's members, the team had access to internal documents and built a deliverable that resonated immediately with the hospital leadership. Future research should verify if our findings are replicable when healthcare teams have different goals, or when they are long established (with a long history of interaction among all members). Our teams had the same goal (i.e. explore challenges and opportunities of a new care delivery system that could result in a quantum leap in improvement of outcomes, patient/family and provider experience), though they varied in team compositions and ties to other units.

Another possible reason for our result on brokerage and performance is connected to a recent research study on social contagion. Centola (2015) built a model of social network formation and demonstrated how breaking down group boundaries to increase the diffusion of knowledge may result in less effective knowledge sharing. Centola's research suggests that complex ideas are more freely integrated across groups if some degree of group boundaries is preserved. This is aligned with the idea that social ties are constrained by individuals' location in social spaces and that their social identities are defined by their participation in social groups (McPherson, 2004; Kossinets and Watts, 2009).

In our study, teams seemed more effective and efficient when fewer cross-ties existed, signaling an increased focus on internal team operation. We also found that the highest-scoring teams used communication media in a parsimonious way. Instead of switching from one communication medium to the other, they chose one or two channels to interact with each other. The most effective teams were able to reduce ambiguity and increase team effectiveness by using only a limited number of channels instead of dispersing time and energy on multiple media (Dennis, Fuller and Valacich, 2008).

## Conclusions and Limitations

This study offers healthcare leaders practical insights on strategies for building teams that are interdisciplinary in nature and have a good balance of external and internal connections. Healthcare leaders would benefit from providing teams with the opportunity to work closely with each other, establishing strong internal connections. In an initial phase, interdisciplinary teams with members representing several medical disciplines and roles need time to brainstorm, learn about individual differences and expertise. The teams in our study were still in the initial stage of the Tuckman's model of team development (Tuckman and Jensen, 1977). After being formed (stage 1), they experience a storming phase (stage 2), where members are more internally focused and are spending time and energy getting to know each other, as well as their potential contribution. That explains why highly performing teams at the time of our study were mostly connected internally rather that with outside members. It would be interesting to explore in another study whether external connections are more prominent in highly effective teams during the final stages of norming and performing, when the team results are planned to be broadcasted to a larger, external audience.

Our study provides some insights to support healthcare decision-makers in their attempt to achieve high value for patients, families and employees (Porter, 2010). We offer empirical evidence to support clinicians and healthcare providers in their attempt to measure outcomes at the institutional and team level using new metrics of knowledge flows and team function. Clinicians are trained to rely only on the "gold standard" of research methodologies, which favor quantitative data and empiricism (Walshe and Rundall, 2001). In this paper we adopted observational methods and qualitative research to inform decision making, providing actionable insights easy to understand an immediately applicable.

While this study seems to confirm the need to favor internal focus which could later be balanced with outward connectivity, it still leaves several questions unanswered, including some raised by Gupta, Smith and Shalley (2006): what is the impact on performance when ideas are being exploited by other individuals or teams? How does organizational politics come into play when it members have to decide what information to share and when they may feel exploited by others?

Our results are based on a sample of teams involved in a specific system redesign project and were composed of a variety of roles including nurses, physicians and medical directors, but also program directors and others holding administrative roles. While their daily activities vary with reference to the immediate impact on patients' health and safety (direct and indirect care), the teams in our study were a good sample of the three types of healthcare teams found in literature: project, management, and care delivery (Lemieux-Charles and McGuire, 2006). A fruitful next step in this research stream would be to replicate the study focusing only on care delivery teams, whose coordination mechanisms and communication processes could be different as they directly involve patients and their families. Future studies could compare teams over a longer period of time as well as teams that are more similar in task, context, and composition (e.g. only nurses and physicians). In that scenario, team performance could be multifaceted to include clinical outcomes, safety events, value, team member ratings of team performance, satisfaction and engagement.

Because of the small sample, we could not implement any statistically significant model to predict team performance by observing knowledge flows, although we got a clear description of a typical scenario in healthcare that could explain differences in performance. Teams who were recognized for their impactful work were engaged in more focused communications within their team and with hospital units and had fewer members who acted as brokering stars. Replicating this study with a larger sample may help establish more support for the theoretical relationships revealed from our study.